\documentclass[aps,twocolumn,preprintnumbers,prd,superscriptaddress,nofootinbib,10pt]{revtex4-2}

\usepackage{amssymb,amsmath,amsthm,amsfonts}
\usepackage[usenames]{color}
\definecolor{darkred}{rgb}{0.5,0,0}

\usepackage{bm}
\usepackage{dcolumn}
\usepackage[utf8]{inputenc}
\usepackage{latexsym}
\usepackage{rotating}
\usepackage{longtable}
\def\arraystretch{1.25}
\usepackage{enumerate}
\usepackage{tensor,multirow}
\usepackage{url}
\usepackage[linktocpage]{hyperref}
\usepackage{float}
\usepackage{graphicx}
\usepackage{xcolor}

\begin{document}

\title{Numerical relativity surrogate waveform models for exotic compact objects: \\ the case of head-on mergers of equal-mass Proca stars}

\author{Raimon Luna $^{*}$}
	\affiliation{Departamento de Astronom\'{i}a y Astrof\'{i}sica, Universitat de Val\`{e}ncia, Dr. Moliner 50, 46100, Burjassot (Val\`{e}ncia), Spain}
        \email{raimonluna@gmail.com}

\author{Miquel Llorens-Monteagudo}
	\affiliation{Departamento de Astronom\'{i}a y Astrof\'{i}sica, Universitat de Val\`{e}ncia, Dr. Moliner 50, 46100, Burjassot (Val\`{e}ncia), Spain}

\author{Ana Lorenzo-Medina $^\dagger$}
	\affiliation{Instituto Galego de F\'{i}sica de Altas Enerx\'{i}as, Universidade de Santiago de Compostela, 15782 Santiago de Compostela, Galicia, Spain}
        \email{analorenzo.medina@usc.es}

\author{Juan Calder\'on~Bustillo $^\ddagger$}
 	\affiliation{Instituto Galego de F\'{i}sica de Altas Enerx\'{i}as, Universidade de Santiago de Compostela, 15782 Santiago de Compostela, Galicia, Spain}
	\affiliation{Department of Physics, The Chinese University of Hong Kong, Shatin, N.T., Hong Kong}
        \email{juan.calderon.bustillo@gmail.com}   
 
\author{Nicolas Sanchis-Gual}
        %\affiliation{Departamento de Matem\'atica da Universidade de Aveiro and CIDMA,
%Campus de Santiago, 3810-183 Aveiro, Portugal}
	\affiliation{Departamento de Astronom\'{i}a y Astrof\'{i}sica, Universitat de Val\`{e}ncia, Dr. Moliner 50, 46100, Burjassot (Val\`{e}ncia), Spain}

\author{Alejandro Torres-Forn\'e}
	\affiliation{Departamento de Astronom\'{i}a y Astrof\'{i}sica, Universitat de Val\`{e}ncia, Dr. Moliner 50, 46100, Burjassot (Val\`{e}ncia), Spain}
        \affiliation{Observatori Astron\`{o}mic, Universitat de Val\`{e}ncia,
C/ Catedr\'{a}tico Jos\'{e} Beltr\'{a}n 2, 46980, Paterna (Val\`{e}ncia), Spain}

\author{Jos\'e A. Font}
	\affiliation{Departamento de Astronom\'{i}a y Astrof\'{i}sica, Universitat de Val\`{e}ncia, Dr. Moliner 50, 46100, Burjassot (Val\`{e}ncia), Spain}
	\affiliation{Observatori Astron\`{o}mic, Universitat de Val\`{e}ncia, C/ Catedr\'{a}tico Jos\'{e} Beltr\'{a}n 2, 46980, Paterna (Val\`{e}ncia), Spain}

\author{Carlos A. R. Herdeiro}
	\affiliation{Departamento de Matemática da Universidade de Aveiro and Centre for Research and Development in Mathematics and Applications (CIDMA), Campus de Santiago, 3810-193 Aveiro, Portugal}

\author{Eugen Radu}
	\affiliation{Departamento de Matemática da Universidade de Aveiro and Centre for Research and Development in Mathematics and Applications (CIDMA), Campus de Santiago, 3810-193 Aveiro, Portugal}

\begin{abstract}
We present several high-accuracy surrogate models for gravitational-wave signals from equal-mass head-on mergers of Proca stars, computed through the Newman-Penrose scalar $\psi_4$. We also discuss the current state of the model extensions to mergers of Proca stars with different masses, and the particular challenges that these present. The models are divided in two main categories: two-stage and monolithic. In the two-stage models, a dimensional reduction algorithm is applied to embed the data in a reduced feature space, which is then interpolated in terms of the physical parameters. For the monolithic models, a single neural network is trained to predict the waveform from the input physical parameter. Our model displays mismatches below $10^{-3}$ with respect to the original numerical waveforms. Finally, we demonstrate the usage of our model in full Bayesian parameter inference through the accurate recovery of numerical relativity signals injected in zero-noise, together with the analysis of GW190521. For the latter, we observe excellent agreement with existing results that make use of full numerical relativity.
\end{abstract}

\maketitle

%%%%%%%%%%%%%%%%%%%%%%%%%%%%%%%%%%%%%%%%%%%%%%%%%%%%%%%%%%%%%%%%%%%%%%%%%%%%%%
\section{Introduction}
\label{Introduction}
%%%%%%%%%%%%%%%%%%%%%%%%%%%%%%%%%%%%%%%%%%%%%%%%%%%%%%%%%%%%%%%%%%%%%%%%%%%%%%

The detection of gravitational waves by the LIGO-Virgo-KAGRA (LVK) detector network has opened a new era of gravitational wave astrophysics \cite{LIGOScientific:2016aoc, LIGOScientific:2017vwq, LIGOScientific:2017ycc, LIGOScientific:2020iuh, LIGOScientific:2018mvr, LIGOScientific:2020ibl, KAGRA:2021vkt,Olsen2022,Nitz2023}. During the first three observing runs these detectors have already detected a large collection of signals, ${\cal O}(100)$, from binary mergers of compact objects, namely black holes and neutron stars, together with an estimation of their physical parameters. These signals have been mainly detected and studied through the matched-filtering of the experimental data \cite{Allen2012,Usman2016,Cannon2021,Chu2022,Veitch2015,Ashton:2018jfp,Zackay2021,Chandra2021,Chandra2022} with theoretical waveform template banks \cite{Babak2006,Cannon2021,gstlal_templatebank}.

Even though templates from the coalescence of black holes and neutron stars have dominated the search landscape as candidates for gravitational wave sources, the possibility of alternative compact objects, known generically as exotic compact objects (ECO), has generated an increasing interest in recent years \cite{Cardoso:2019rvt}. In particular, much scientific attention has been oriented towards those objects that could generate gravitational-wave signals resembling those of black holes, which could generate confusion in their classification. These objects are typically known as ``black hole mimickers''.

A particularly important class of ECOs are those based on gravitationally bound states of ultralight bosonic fields \cite{Kaup:1968zz, Ruffini:1969qy}. These fields could appear in particle physics models, such as the string axiverse \cite{Arvanitaki:2009fg, Arvanitaki:2010sy} or as extensions of the Standard Model \cite{Freitas:2021cfi}. Ultralight bosonic fields are capable of forming stationary states, resembling hydrogen orbitals \cite{Herdeiro:2020kvf}, which can also be rotating \cite{schunck1996rotating, Herdeiro:2019mbz}. In the case of massive complex vector fields --- Proca fields --- the objects are known as Proca stars \cite{Brito:2015pxa}.

While sharing some similarities, scalar and Proca bosonic stars have distinct behaviors regarding their formation and stability \cite{Sanchis-Gual:2019ljs}. In particular, some configurations of rotating bosonic stars exhibit a bar-mode instability \cite{DiGiovanni:2020ror} which can be prevented through the addition of non-linear interaction terms in their action \cite{Siemonsen:2020hcg,dmitriev2021instability} or considering multi-state bosonic configurations~\cite{Sanchis-Gual:2021edp}. On the other hand, Proca stars are known to have sufficiently generic mechanisms of formation  though the so-called  ``gravitational cooling'' mechanism \cite{Seidel:1993zk, DiGiovanni:2018bvo} and a stable ground state \cite{Herdeiro:2023wqf}. A general review on boson stars can be found in \cite{Liebling:2012fv}.

Numerical evolutions of Proca stars have been performed at the fully non-linear regime in 3+1 numerical relativity, both with single stars \cite{Sanchis-Gual:2017bhw} and head-on collisions \cite{Sanchis-Gual:2018oui, Sanchis-Gual:2022mkk}. This has allowed the use of the gravitational waveforms extracted from the simulations to compare with actual detections and perform parameter estimations (PE) \cite{CalderonBustillo:2022dph, CalderonBustillo:2022cja}. In particular, statistical inference on the event GW190521 resulted in a very good agreement, even with a slightly higher Bayes factor than with black hole templates \cite{CalderonBustillo:2020fyi}.

The standard procedure for Bayesian PE involve sampling over a large number of waveform templates generated with different physical parameters, typically on the order of millions. As numerical relativity simulations are computationally expensive, on the order of thousands of CPU-hours per simulation, generating all the templates by direct simulation is not possible at the practical level. For the specific case of binary black hole (BBH) mergers, the development of waveform templates (or ``approximants'') based upon approximations to the equations of motion of the two-body-problem in general relativity, has already a long history spanning over two decades \cite{Schmidt2020_review}. For the three main model families covering the whole inspiral-merger-ringdown process, namely \texttt{SEOBNR} \cite{Buonanno:1998gg,Buonanno:2000ef,SEOBNRv4,SEOBNRv4PHM}, \texttt{TEOBResumS} \cite{Nagar2018,Nagar2023,Gonzalez2023} and \texttt{IMRPhenom} \cite{Ajith:2007qp,Santamaria:2010yb,PhenomPv3HM,XPHM_Pratten},  waveform templates are either calibrated to numerical relativity using analytical or semi-analytical expressions or they are obtained through the ``hybridization'' of waveforms from numerical relativity simulations and post-Newtonian approximations.

A fourth and more recent line of work is the construction of {\it surrogate} waveform models, first proposed in the seminal work of~\cite{Field:2014}. Such models have been shown to reach an accuracy comparable to numerical relativity and they will be the focus of our work (albeit in the context of Proca-star mergers). For non-spinning quasi-circular BBH mergers up to a mass-ratio $q=8$, the first attempt to build a surrogate model employing waveforms from numerical relativity was performed by~\cite{Blackman:2015}. This work was soon followed by the first surrogate model for precessing systems~\cite{Blackman:2017a}, \texttt{NRSur4ds2}, built from a set of 276 precessing BBH simulations, and by the corresponding extensions to account for unequal masses and generic spins \texttt{NRSur7dq2}~\cite{Blackman:2017b} and \texttt{NRSur7dq4}~\cite{Varma:2019}. The latter was calibrated with 1528 numerical relativity precessing simulations with mass-ratio $q\leq 4$ and spin magnitudes up to 0.8, covering all harmonics with $l\leq 4$, retaining all intrinsic degrees of freedom of quasi-circular BBH systems. \texttt{NRSur7dq4} is the most accurate waveform model for quasi-circular BBH with generic spins. This model has been employed to analyze several exceptional events from the LVK Collaboration~\cite{GW190412,LIGOScientific:2020iuh}, to extract measurements of novel observables like orbital precession \cite{Hannam_nature_precession}, gravitational-wave recoils \cite{Varma2022_kick,Kick_GW190412} or the Chern-Pontryagin pseudo-scalar \cite{AdriNico}. 

In addition, Islam et al.~\cite{NRSurCatalog} recently used \texttt{NRSur7dq4} to re-analyze 42 events detected by the LVK Collaboration. Moreover, for aligned-spin binaries, a surrogate model for hybridized numerical relativity waveforms was constructed in~\cite{Varma:2019b} using 104 simulations, employing Effective-One-Body for hybridizing the orbital frequency  and post-Newtonian results for hybridizing the harmonic amplitudes. This was extended in~\cite{Yoo:2022} who developed a non-precessing 2-dimensional surrogate model of hybridized waveforms up to mass-ratio $q=15$, \texttt{NRHybSur2dq15}, in order to build a template waveform for signal GW190814~\cite{GW190814}. It is also worth mentioning that efforts based on Machine Learning, namely Gaussian Process Regression, have been used to build waveform surrogate models for both non-precessing and precessing BBH systems at points of the parameter space not covered by numerical relativity simulations~\cite{Williams:2020, Andrade:2023sal}. 

The progress in surrogate waveform models for BBH mergers motivates the development of such models for ECOs, in particular if a systematic search for Proca stars is to be pursued. These models should be able to produce accurate waveform estimates, taking as input the physical parameters of the system, using modest computational resources. Some attempts have been made in this direction by application of generative adversarial networks (GAN)~\cite{Freitas:2022xvg}, already producing promising results. In this paper, we perform a first approach to the use of surrogate models for PE in Proca star binaries by restricting to head-on collisions of Proca stars with equal masses. The choice of head-on collisions instead of quasi-circular inspirals is entirely motivated by the much higher technical difficulty and computational cost of the Proca inspirals, which makes it challenging to perform numerical-relativity simulations of such systems at the present time.

Therefore, in our case, the system depends on a single physical parameter, the normalised frequency of the system $\omega/\mu_V$,  where $\mu_V$ denotes the mass of the ultralight vector boson forming the stars. For the PE procedure to confidently resolve the physical parameters of the system, the error introduced by the surrogate model must be much lower than the typical differences between templates in the sampling range. For this reason, we analyze a variety of possible surrogate architectures that can be used to fit the simulation data. We compare their performances on a testing set and we finally select the most reliable one for the actual Bayesian analysis. The models in this study are data-driven, based entirely on a catalog of simulations \cite{Sanchis-Gual:2022mkk}, with no built-in physical information.

In the first class of models, which we call two-phase architectures, we encode the waveform data into a feature space of lower dimension. This drastically reduces the amount of numerical parameters that need to be fitted, thus reducing the model size. The parameters in the feature space (sometimes called latent space) can then be interpolated in terms of $\omega/\mu_V$. As we have only one independent variable, we use standard cubic spline interpolation. The algorithms that we analyze for the dimensional reduction phase are singular value decomposition, empirical interpolant representation \cite{Field:2013cfa} and a deep convolutional autoencoder. The first two algorithms decompose the data as a linear combination of a suitable reduced basis, while the last one uses a nonlinear mapping based on neural networks. The second type of model is a more straightforward approach, which we call monolithic, as a single neural network is used to map the frequency values $\omega/\mu_V$ to the temporal values of the waveform.

We use our model to perform Bayesian parameter inference on both simulated and real data. In the first case, we perform parameter inference on numerical relativity waveforms injected in zero noise, demonstrating that our surrogate can accurately estimate the parameters of such injections. In the second case, we perform parameter inference on the data containing the signal GW190521, showing that our surrogates faithfully match those obtained through numerical relativity waveforms, presented in \cite{CalderonBustillo:2020fyi,CalderonBustillo:2022dph}, while using significantly less computational resources. This indicates that this type of data-driven surrogate models are a very promising tool to encapsulate the catalogs of numerical boson-star waveforms in order to be efficiently used in detection data analysis.

The paper is structured as follows. In Section \ref{sec:dataset} we present the generation of the training data and its preparation before it can be fed to the surrogate models. We also discuss the criteria used for the evaluation and comparison of the different models. Section \ref{sec:Two-stage} describes the different architectures of two-stage models, i.e., those who can be divided into encoder and translator modules. A different type of models, the monolithic surrogates, are exposed in Section \ref{sec:Monolithic}. The results of model evaluations, as well as the Bayesian parameter estimations performed with the models, are discussed in Section \ref{sec:Results}. We conclude in Section \ref{sec:Discussion}.
%
%%%%%%%%%%%%%%%%%%%%%%%%%%%%%%%%%%%%%%%%%%%%%%%%%%%%%%%%%%%%%%%%%%%%%%%%%%%%%%
\section{The Dataset}
\label{sec:dataset}
%%%%%%%%%%%%%%%%%%%%%%%%%%%%%%%%%%%%%%%%%%%%%%%%%%%%%%%%%%%%%%%%%%%%%%%%%%%%%%

\subsection{The Proca Merger Catalog}

The models are trained on a catalog of 59 equal-mass simulations from \cite{Sanchis-Gual:2022mkk}, of Proca stars in Einstein-Proca theory
\begin{equation}
    \mathcal{S} = \int d^4 x \sqrt{-g} \left(\frac{R}{16\pi} - \frac 14 F_{\mu\nu} \bar F^{\mu\nu} - \frac 12 \mu^2 A_\mu \bar A^\mu  \right) \; ,
\end{equation}
where $A_{\mu}$ and $F_{\mu\nu}$ are the Proca potential and field strength, respectively, $\mu$ is the Proca field mass parameter, and the bar denotes complex conjugation. We also take $c = G = 1$. The interested reader is addressed to \cite{Sanchis-Gual:2022mkk} and references therein for details on the formalism. The simulations were performed in the \texttt{Einstein Toolkit} \cite{EinsteinToolkit:2021_11}, which uses the \texttt{Cactus} framework \cite{Goodale:2002a}, under the Baumgarte-Shapiro-Shibata-Nakamura (BSSN) formulation using the \texttt{McLachlan} \cite{Brown:2008sb, Reisswig:2010cd} and \texttt{Lean}~\cite{ZilhaoWitekCanudaRepository} thorns. Details on the simulations and construction of initial data can be found in \cite{Brito:2015pxa, Sanchis-Gual:2017bhw, Sanchis-Gual:2018oui}. The thorn containing the Proca equations was implemented in \cite{Zilhao:2015tya} and is available in~\cite{Canuda_2023}. 

Each simulation produces gravitational waveforms for the Newman-Penrose scalar $\psi_4$, which are decomposed in spherical harmonic modes. Our study will be centered on the fundamental $(l,m) = (2,2)$ and $(l,m) = (2,0)$ modes. An example of the waveforms, together with their prediction by one of the surrogates presented below, is shown in 
Figure~\ref{fig:Waveform_Example}.
\begin{figure}[thpb]
\begin{center}
\includegraphics[width=0.45\textwidth]{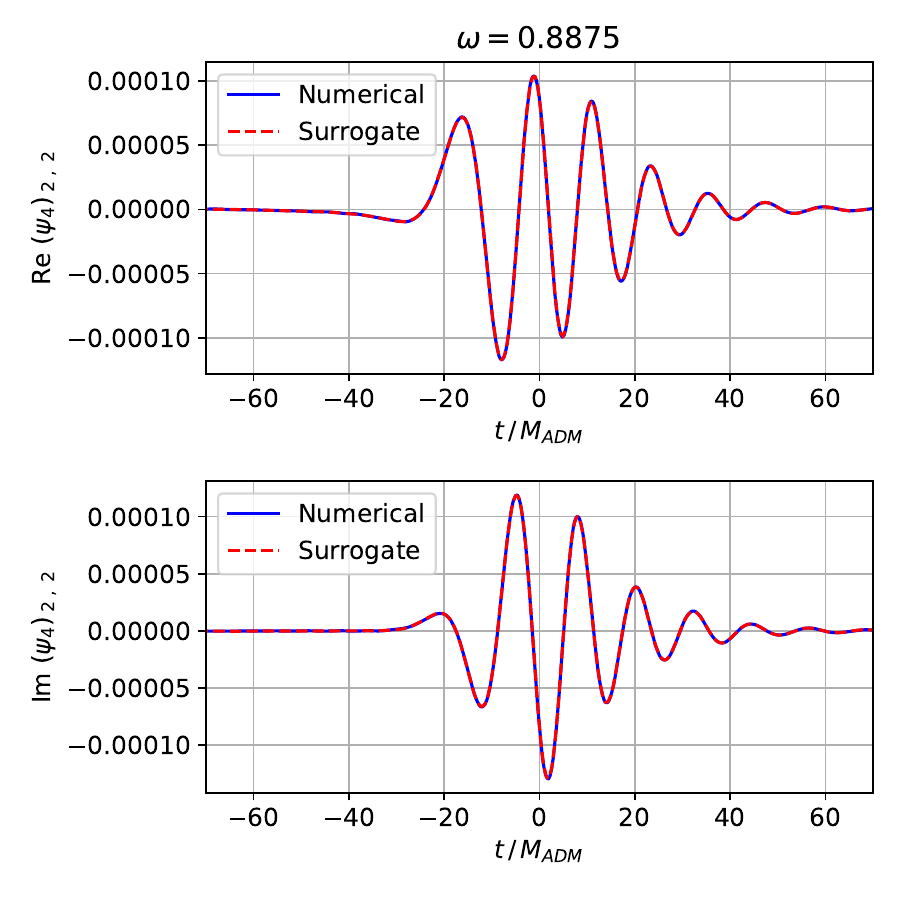}
\caption{Example of the reconstruction of a $\psi_4$ waveform by the SVD surrogate model, for the real (top) and imaginary (bottom) parts of the mode $(l,m) = (2,2)$ at $\omega = 0.8875$. \label{fig:Waveform_Example}}
\end{center}
\end{figure}
Even though the ADM mass of the initial data may vary due to the construction of the Proca star binaries, the numerical relativity waveforms are rescaled so that $M_{\rm{ADM}} = 1$ before they are treated in the surrogate pipelines.
\subsection{Pre-Processing of the Data}
\label{subsec:pre-processing}
\begin{figure}[thpb]
\begin{center}
\includegraphics[width=0.45\textwidth]{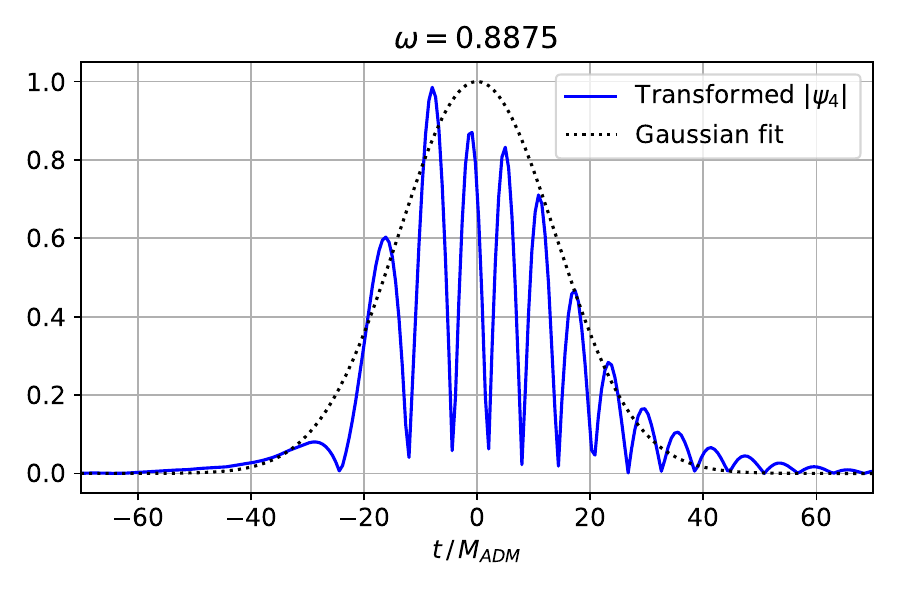}
\caption{Transformed profile for the amplitude $|\psi_4|$ of the mode $(l,m) = (2,2)$ at $\omega = 0.8875$, together with the fitted Gaussian curve. After the transformation, the peak of the Gaussian lies at coordinates $(x, y) = (0, 1)$.  \label{fig:Gaussian_Fit}}
\end{center}
\end{figure}

The complex $\psi_4$ waveforms of the original dataset are subjected to a pre-processing pipeline before they are submitted to the actual model. There are a number of reasons for this. Firstly, the algorithms will always tend to be more effective when the samples are similar to each other. For instance, having large time shifts between otherwise very similar waveforms can unnecessarily increase the cost of the fitting. Secondly, many machine learning algorithms work best with data of numerical value of order one. As the values of $\psi_4$ vary usually on the order of $10^{-4}$, it is highly advisable to rescale the data before making any learning attempt on it. Additionally, numerical relativity simulations usually include a burst of unphysical junk radiation at the beginning of the waveform, which has to be removed. And last but not least, it is highly convenient to have all waveforms on the same time grid, which makes it possible to treat them on equal footing. To this end, a Gaussian curve is fitted to the modulus $|\psi_4|$ of the waves, as
\begin{equation}
    f(t) = A e^{- (t - t_0)^2/s^2}\; .
\end{equation}
Then, the waveforms are time-shifted by $t_0$ so that all the waves are centered at $t = 0$. Then, they are divided by $A$ so their range of oscillation becomes of order one. In order to perform the transformations, the original waveform is interpolated by cubic splines and re-evaluated on a fixed grid of $N_\psi$ equally spaced points between -150 and 150. $\psi_4$ is set to zero for $t < 100$, to remove the junk radiation. Figure  \ref{fig:Gaussian_Fit} shows an example of the Gaussian curve fit to the amplitude of a wave after the transformations. 

In order for the surrogate to be used for parameter estimation, the relative time positions of the different modes have to be preserved. The values of the time shift and amplitude $(t_0, A)$ are stored in the dataset, as they also need to be interpolated in order to recover the information later. Once the model is evaluated, we multiply the waveforms by their respective amplitudes $A$ and we time-shift them so that the peak of the reference mode $(l,m) = (2,2)$ lies at $t = 0$. This way, the original relative time positions of all the modes are preserved.

The method of the Gaussian fit for the identification of the peak and amplitude of the waveform, as opposed to just taking the absolute maximum of $|\psi_4|$, guarantees that the paramters $A$ and $t_0$ will vary continuously with $\omega/\mu_V$. The absolute maximum of $|\psi_4|$, on the other hand, can suffer discontinuous jumps between relative maxima, which would create discontinuous jumps on the data, thus damaging the performance of the fitting algorithms.
\subsection{Model Testing}
\begin{figure}[thpb]
\begin{center}
\includegraphics[width=0.45\textwidth]{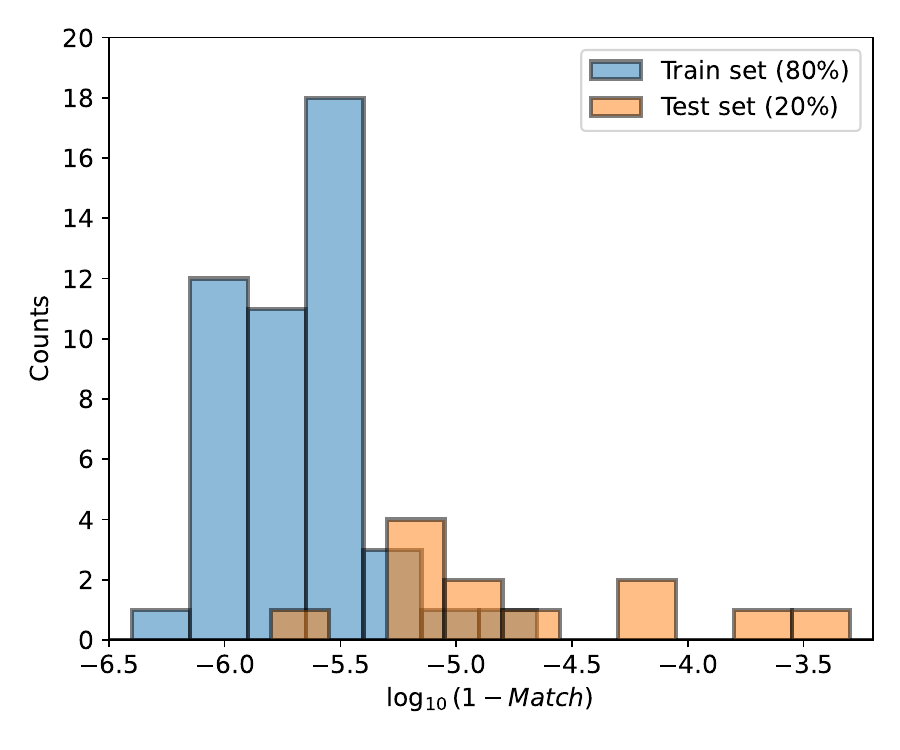}
\caption{Histograms of mismatches for the training and testing sets of the SVD surrogate model for the mode $(l,m) = (2,2)$. Some amount of overfitting can be appreciated, as the model performs worse on the testing data (which has not been used in the training), than on the training data itself. \label{fig:Train_Test_Hist}}
\end{center}
\end{figure}

As in any data-driven learning algorithm, the surrogate models are prone to suffer from overfitting. In other words, the models will tend to perform better on the data they have been trained than on new data that they have not seen before. An example of this is shown in Fig.~\ref{fig:Train_Test_Hist}. Therefore, before the surrogate model can be used on parameter estimation samplers, an estimate must be done of its reliability when generating waveforms outside of its own training dataset. 

The full waveform dataset from the original Proca catalog is divided into training and testing datasets with 80\% and 20\% of the samples, respectively. All the model fittings will be performed on the training dataset, and the test set will be reserved as a final evaluation of the model. As a metric for the accuracy of the model we employ the mismatch \cite{Cutler:1994ys, Apostolatos:1995pj, Finn:1992wt} between the predicted and the actual waveform, defined as $1 - M(\psi_1, \psi_2)$. Here, the match $M(\psi_1, \psi_2)$ is given by
\begin{equation}
  M(\psi_1,\psi_2) = \max_{t_s,\phi}\frac{\langle \psi_1 | \psi_2 \rangle}{\sqrt{\langle \psi_1 | \psi_1 \rangle \langle \psi_2 | \psi_2 \rangle}},
\end{equation}
with $\langle \psi_1 | \psi_2 \rangle$ the inner product defined as
\begin{equation}
  \langle \psi_1 | \psi_2 \rangle = 4 \times {\rm Re}\int_{\nu_0}^{\nu_{\rm 1}} \frac{\tilde{\psi_1}(\nu) \, \tilde{\psi_2}^{*}(\nu)}{S_n(\nu)}\,d\nu\,,
\end{equation}
where $\tilde \psi_i(\nu)$ are the Fourier transforms, and $\tilde \psi_i^*(\nu)$ their complex conjugates. The match is maximized over relative time shifts $t_s$ and phases $\phi$ between $\psi_1$ and $\psi_2$. $S_n(f)$ is the one-sided noise power spectral density (PSD), which we take to be flat.
\subsection{Model Selection}
Before the final evaluation on the test set, we have to choose among a number of different architectures of models, as well as the hyperparameters intrinsic to each architecture. It is not convenient, however, to check the performance of each configuration on the final test set. If we did so, we could again be selecting the architecture that best suits our particular test set, without it generalizing to arbitrary data.

While this could be easily solved by splitting the training set into two subsets ---one for yet again training and the other for validation--- it would further reduce the number of samples in each subset. Given the limited number of simulations in our catalog, it would be convenient to use a scheme where we could extract information from all the data in the original training set. A good solution for this is $k$-Fold cross-validation on the training set. In this scheme, the training set is itself divided in $k$ (approximately) equal parts, called folds. We then train $k$ versions of the model. For each version, we train the model on $k-1$ of the folds, and we validate it on the remaining fold. With this method, all of the training data samples have been used as a validation of the model architecture, without ever being used both in the training and validation sets. It is important to keep in mind that the final testing set does not intervene at all during the model selection phase. In this paper we will always use $k=5$.

\begin{figure}[t]
\begin{center}
\includegraphics[width=0.45\textwidth]{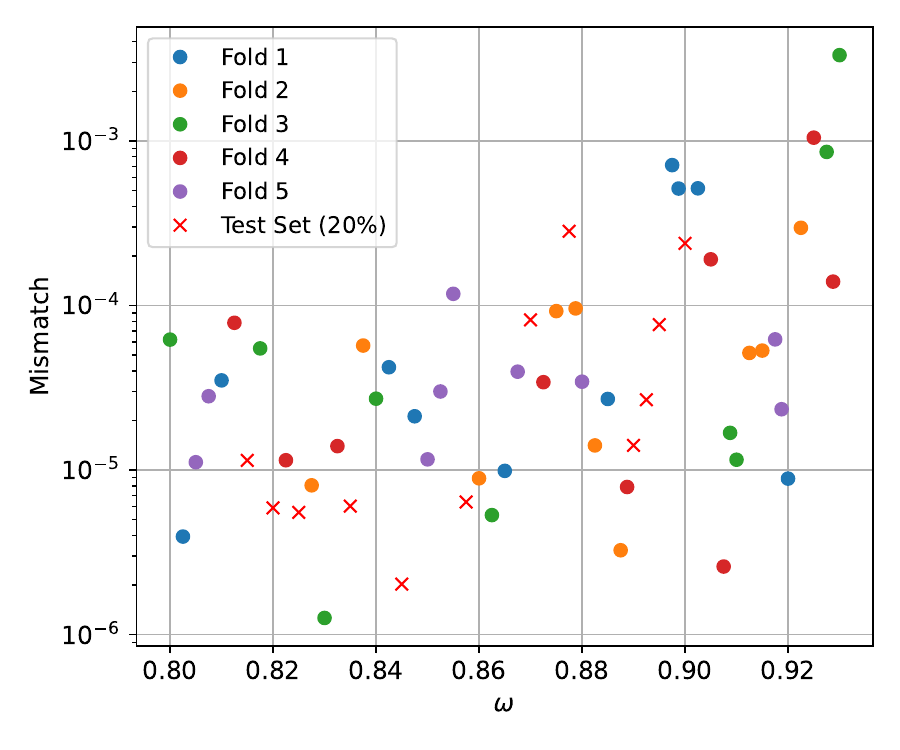}
\caption{Mismatches of the empirical interpolant representation model on the dataset, for the mode $(l,m) = (2,2)$. For each of the 5 folds in the training set, shown by circular dots, the model has been trained on the other 4 folds. For the test set, shown by crosses, the model has been trained on the 5 folds in the training set. \label{fig:Folds_Test_Scatter}}
\end{center}
\end{figure}

In Fig.~\ref{fig:Folds_Test_Scatter} we depict the mismatches of all the $(l,m) = (2,2)$ waveforms in the dataset, evaluated on their corresponding model, in this case based on the empirical interpolant representation (see Sections \ref{subsec:SVD} and \ref{subsec:EIR}). The waveforms of the training set, which we depict as circular dots, are divided in the 5 cross-validation folds, corresponding to the 5 different colors of the dots. For the validation on each of the folds, the model has been trained on the remaining 4 folds. On the other hand, for the waveforms in the final test set, the model has been trained on the full training set (all 5 folds together). The distribution of mismatches is quite similar for all the subsets, which indicates that our statistical estimates of the surrogate performance are consistent with each other. The points at the edge of the domain, close to $\omega = 0.93$, have significantly higher mismatches as extrapolation is being used.

For the sake of consistency, in all of the dataset splittings, we keep the different modes of each simulation together. In other words, if the mode $(l,m) = (2,2)$ is in one particular subset, then so are the other values of $(l,m)$.
%
%%%%%%%%%%%%%%%%%%%%%%%%%%%%%%%%%%%%%%%%%%%%%%%%%%%%%%%%%%%%%%%%%%%%%%%%%%%%%%
\section{Two-Stage Models}
\label{sec:Two-stage}
%%%%%%%%%%%%%%%%%%%%%%%%%%%%%%%%%%%%%%%%%%%%%%%%%%%%%%%%%%%%%%%%%%%%%%%%%%%%%%
%
The two-stage surrogate models are divided in a pipeline of two processes, that we chose to name as encoder-decoder and translator. 
\begin{itemize}
    \item The encoder-decoder phase is an invertible process of dimensional reduction from the $N_\psi$ complex values of $\psi_4^i$ to a feature space of lower dimension $N_Z$, which can be complex or real depending on the algorithm.
    \item The translator phase interpolates the features $Z^j$, as well as the amplitude and peak position, as a function of the physical parameters of the model. In this case, the only physical parameter is the frequency $\omega/\mu_V$ of the merging Proca stars.
\end{itemize}

Schematically,
\begin{equation}
\begin{split}
    \mathbb{R} &\rightarrow \mathbb{C}^{N_Z} \rightarrow \mathbb{C}^{N_{\psi}} \\
    \omega &\mapsto Z^j \mapsto \psi_4^i \; .
\end{split}
\end{equation}
When the model is evaluated for parameter estimation, the process is then reversed. First, the physical frequencies are used in the translator to obtain the coordinates $Z^j$ in feature space, which then are passed through the decoder to recover the values of the Newman-Penrose scalar $\psi_4$. 
\subsection{Singular Value Decomposition}
\label{subsec:SVD}
Our first approach for dimensional reduction is singular value decomposition (SVD). This produces an orthonormal complex basis for our space of waveforms, in such a way that successive basis vectors are less and less important in the accurate description of the data. The notion of importance is quantitatively encoded in the singular values of the basis. Namely, if we construct an $N_\psi \times N_t$ matrix $M$, whose $N_t$ columns are the training waveforms, we can decompose it as
\begin{equation}
    M = U \Sigma V^\dagger\; .
\end{equation}
Here the columns of $U$ are the adapted orthonormal basis $u_i$, and the diagonal elements of $\Sigma$ are their corresponding singular values $\sigma_i$. By taking the first $N_Z$ of such vectors, we can express a compressed version of any waveform as a linear combination of them. Their $N_Z$ coordinates define the feature space of the model. The fact that $U$ is a unitary matrix greatly simplifies its inversion, as $U^{-1} = U^\dagger$, so the encoding-decoding of waveforms becomes trivial.

\begin{figure}[t]
\begin{center}
\includegraphics[width=0.45\textwidth]{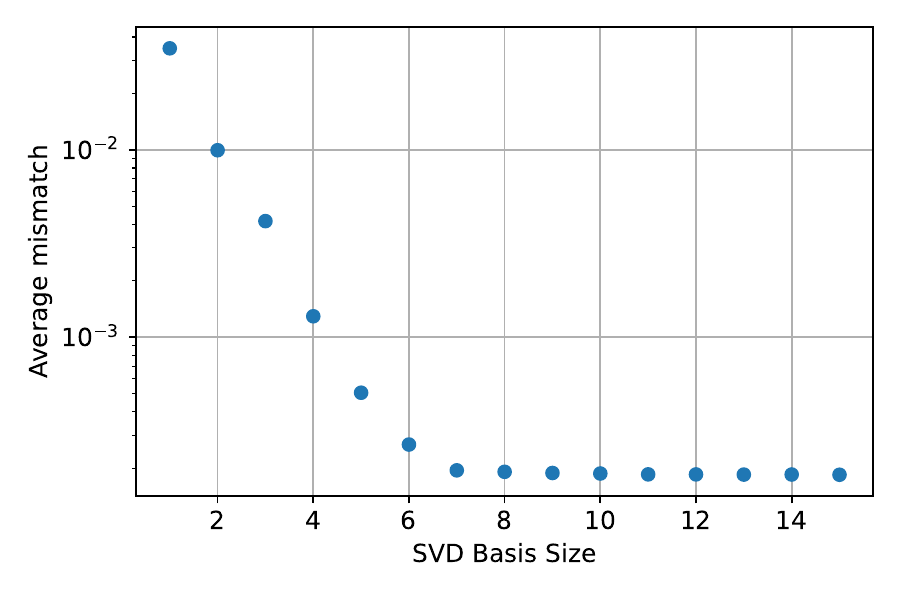}
\caption{Average mismatch of the SVD model on the $k$-Fold cross-validation of the test set, for the mode $(l,m) = (2,2)$. The mismatch decreases exponentially with the number $N_Z$ of vectors until $N_Z=7$. Adding more vectors does not seem to help.  \label{fig:SVD_Basis_Size}}
\end{center}
\end{figure}

The main hyperparameter on the SVD models is the number of vectors in the reduced basis $N_Z$. We have thus performed an analysis of such basis size on the performance of the model, as shown in Fig.~\ref{fig:SVD_Basis_Size}. The averaged mismatch after the $k$-Fold cross validation process decays approximately as an exponential law for $1 \leq N_Z \leq 7$, and remains fairly constant after that. In other words, having more than 7 vectors in the reduced basis does not seem to increase the precision of the model. We have performed our parameter estimation with a basis of 10 vectors, to keep a safe margin.
\subsection{Empirical Interpolant Representation}
\label{subsec:EIR}
The empirical interpolant representation (EIR) is an additional step  we take after SVD to obtain a more convenient basis for the space of waveforms in our model. The idea behind EIR algorithms is to choose a reduced basis $B_j$ of the waveform space and a subset of indices 
\begin{equation}
\mathcal{I} \subset \mathbb{Z}_{N_\psi} \, , \quad |\mathcal{I}| = N_Z
\end{equation}
such that 
\begin{equation}
    B^i_j = \delta^i_j \quad \forall j \in \mathcal{I} \; .
\end{equation}
In other words, the coordinates of $\psi_4(t)$ in the basis $B_j$ are its values at certain selected times $\psi_4(t = T^j) = \psi_4^j$, for $j \subset \mathcal{I}$, as
\begin{equation}
    \psi_4^i = \sum_{j \in \mathcal{I}} \psi_4^j \, B^i_j\; . 
\end{equation}
This process is performed by the Python library \texttt{rompy} \cite{Field:2013cfa, rompy-2020}, which takes as an input the SVD basis $u_i\, ,(i = 1, \dots, N_Z)$ and computes the basis vectors $B_j$ as well as the relevant indices $\mathcal{I}$. 

The encoding procedure of the waveforms into the reduced feature space is then straightforward, as it just involves taking the elements of $\psi_4$ in the selected positions $\mathcal{I}$. The decoding is also simple, just multiplying the feature space coordinates by the basis vectors $B_j$.

\begin{figure}[t]
\begin{center}
\includegraphics[width=0.45\textwidth]{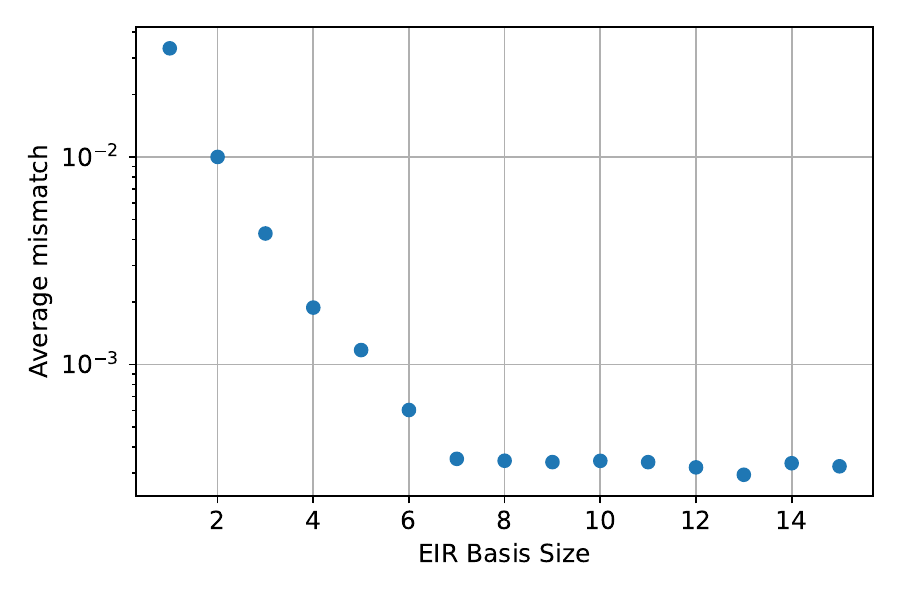}
\caption{Average mismatch of the EIR model on the $k$-Fold cross-validation of the test set, for the mode $(l,m) = (2,2)$. The mismatch decreases exponentially with the number $N_Z$ of vectors until $N_Z=7$. Adding more vectors does not seem to help.  \label{fig:EIR_Basis_Size}}
\end{center}
\end{figure}

In a similar way as in the case of SVD, we have studied the mismatch dependence on the number of EIR basis vectors $N_Z$, as shown in Fig.~\ref{fig:EIR_Basis_Size}. The plot looks very similar to that of SVD, with an exponential decay which stabilizes for $N_z \geq 7$.
\subsection{Deep Convolutional Autoencoder}
Autoencoders provide a nonlinear alternative to EIR dimensional reduction. In this case, we set up two 1D convolutional neural networks, which will act as an encoder and a decoder, respectively. The encoder will learn a map from the real and imaginary parts of the waveform $\psi_4^i$ to the feature space $Z^j$, which is now taken to be real. In other words, the encoder will become a nonlinear mapping from $\mathbb{C}^{N_\psi}$ to $\mathbb{R}^{N_Z}$. On the other hand, the decoder will learn the inverse mapping, trying to recover $\psi_4^i$ from the features $Z^j$. In the context of autoencoders, the reduced feature space $Z^j$ is typically known as latent space.
\begin{figure}[thpb]
\begin{center}
\includegraphics[width=0.45\textwidth]{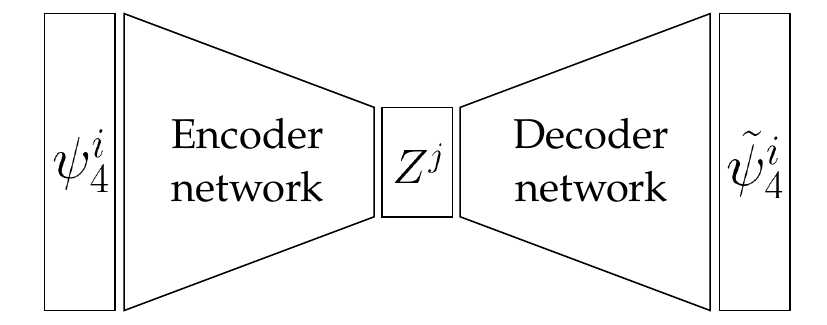}
\caption{Schematic diagram of the autoencoder architecture. Two deep convolutional networks, the encoder and the decoder, have to reconstruct the original signal by encoding the information in a reduced latent space. Once the autoencoder is trained, the decoder can be used on its own to produce waveforms from vectors of coordinates $Z^j$ in the latent space. \label{fig:AutoencoderScheme}}
\end{center}
\end{figure}

\begin{figure}[t]
\begin{center}
\includegraphics[width=0.45\textwidth]{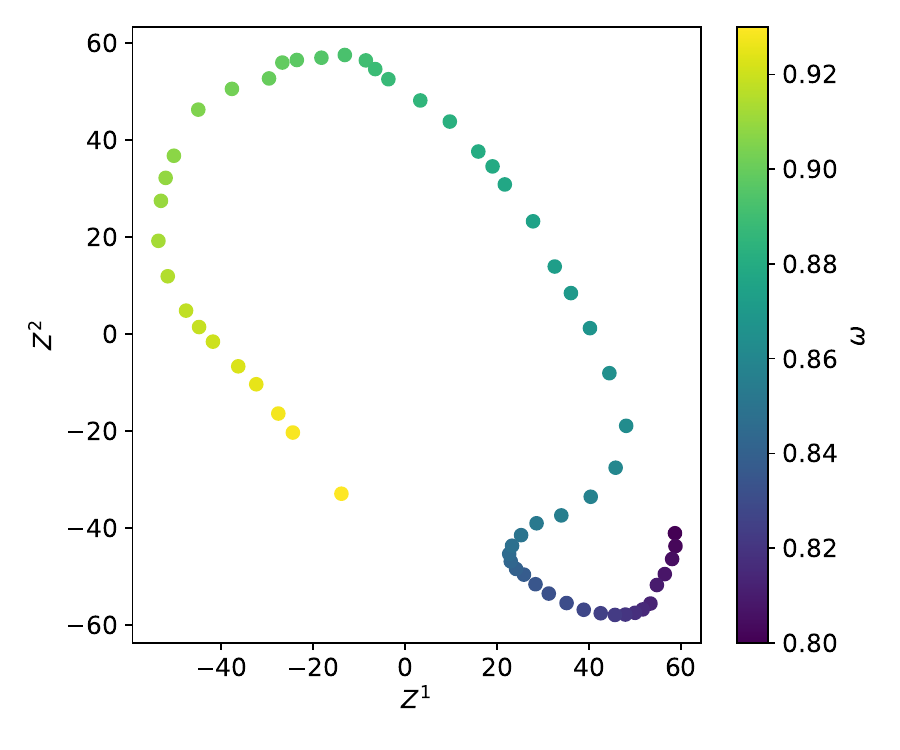}
\caption{Distribution of the dataset waveforms in the 2-dimensional latent space, with coordinates $(Z^1, Z^2)$, for the mode $(l,m) = (2,2)$. The autoencoder has autonomously ordered the waveforms by their governing physical parameter $\omega/\mu_V$ along a continuous curve embedded in the latent space.  \label{fig:LatentSpaceMap}}
\end{center}
\end{figure}

The training process is performed on a concatenation of the encoder and decoder networks, in such a way that the goal of the full system is to learn the identity function, with the output being as close as possible to the input. This process is nontrivial as the number of parameters has to be drastically reduced from $N_\psi$ to $N_Z$ as the information is transmitted through the bottleneck between the networks. A schematic representation of this configuration is shown in 
Fig.~\ref{fig:AutoencoderScheme}. We define the loss function of the system as
\begin{equation}
    \mathcal{L} = \frac{1}{2N_\psi} \sum_{i=1}^{N_\psi} |\psi_4^i - \tilde \psi_4^i|^2\, ,
\end{equation}
with $\psi_4^i$ and $\tilde \psi_4^i$ the original and recovered waveforms, respectively. The loss is also averaged over the instances in the training batch. After the training, the encoder and decoder modules can be used separately to map the waveforms to their coordinates in the latent space and vice-versa. Figure \ref{fig:LatentSpaceMap} shows the embedding of the waveforms in a latent space of $N_Z=2$, where the model has spontaneously ordered the waveforms by their value of $\omega/\mu_V$. In this case the mapping is nonlinear, so the number of coordinates in the latent space is usually lower than the number of basis vectors that are needed in SVD or EIR encoding.

The structure of the networks is fairly symmetric between encoder-decoder. The encoder receives the training minibatch of waveforms as a tensor of dimensions $(N_B, 2, N_\psi)$. $N_B = 16$ is the number of instances in the training minibatch, 2 is the number of channels (corresponding to the real and imaginary parts of the wave), and $N_\psi = 512$ is the length of the time series containing the wave. The batch is first passed through a 1-dimensional convolutional layer of 16 filters of kernel size 4, stride 2 and padding 1. After that a leaky ReLu activation function, with a negative slope of 0.2, is applied.  Then the batch goes through a series of 6 more convolutions with identical kernel configurations, but each time duplicating the number of filters. After each convolution, we apply a batch normalization and another leaky ReLu with a negative slope of 0.2. This configuration of filters reduces the length of the data samples by 2 at each convolution, while doubling the number of channels. Finally a last convolution is applied, this time with a stride of 1 and padding reduced to 0, with $N_Z$ filters. This reduces the minibatch to a tensor of size $(N_B, N_Z, 1)$, where the $N_Z$ channel values are taken to be the coordinates in the latent space. 

The architecture of the decoder is quite exactly inverse from the encoder. The input minibatch of dimensions $(N_B, N_Z, 1)$ is passed through a transposed convolution with 64 filters of kernel size 4, stride 1 and padding 0. Then, a series of 7 more (transposed) convolutions are applied, now with a stride of 2 and padding of 1, each one duplicating the size of the data instances and halving the number of filters. After each convolution (except the very last one), we apply batch normalization and a ReLU activation. The last convolution actually reduces the number of filters down to 2 (real and imaginary parts of $\psi_4$), resulting in an output of dimensions $(N_B, 2, N_\psi)$, as expected.
\begin{figure}[t]
\begin{center}
\includegraphics[width=0.45\textwidth]{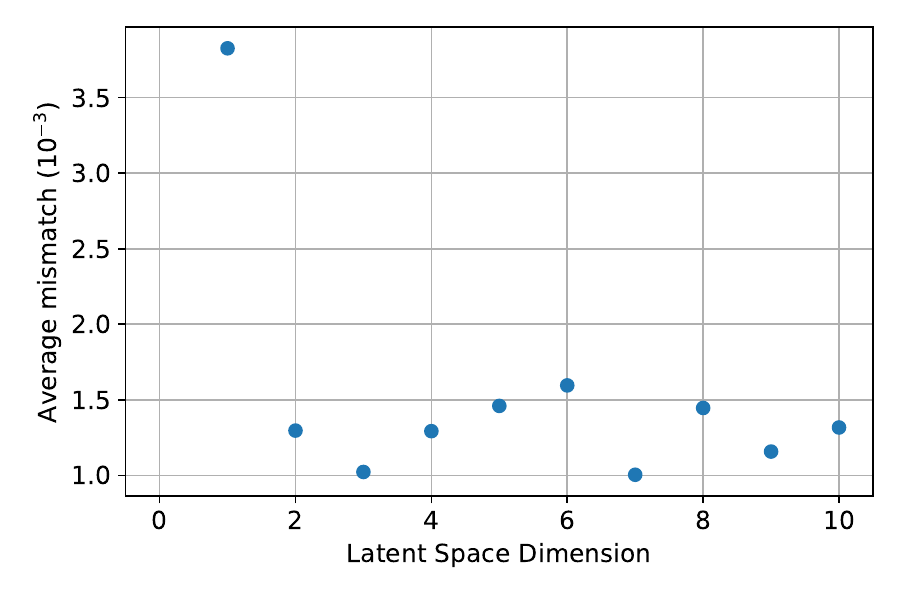}
\caption{Average mismatch of the autoencoder model on the $k$-Fold cross-validation of the test set, for the mode $(l,m) = (2,2)$. The performance does not seem to vary as long as the latent space dimension $N_Z$ is strictly larger than 1. \label{fig:Latent_Space_Dimension}}
\end{center}
\end{figure}

The training is performed on a Nvidia RTX 4090 24GB GPU over minibatches of 16 waveforms going through 500 iteration epochs with a learning rate of $10^{-3}$, using an Adam optimizer with parameters $(\beta_1, \beta_2) = (0.5, 0.999)$, in the framework \texttt{PyTorch} \cite{NEURIPS2019_9015}. 

Figure \ref{fig:Latent_Space_Dimension} shows the average mismatch of the surrogate as a function of the latent space dimension $N_Z$, computed over the $k$-Fold cross validation process. Even though the waveforms depend on a single physical parameter, namely the frequency $\omega/\mu_V$, the model has proven to perform much better if we allow at least 2 dimensions in the latent space. A latent dimension higher than that does not have a large influence on the mismatch. For this reason, the final selected model has ben chosen to have $N_Z = 2$.

\subsection{The Translator Phase}

The task of the translator phase of the model is to interpolate the coordinates in feature space $F^j$ as a function of the Proca star frequency $\omega/\mu_V$. As there is only one independent variable, order 3 spline interpolation is an efficient and fast method. These are provided by the \texttt{scipy.interpolate} function interp1d, which is applied to every coordinate in the latent space, as well as the peak position and amplitudes $(t_0, A)$, as a function of the variable $\omega/\mu_V$.
%
%%%%%%%%%%%%%%%%%%%%%%%%%%%%%%%%%%%%%%%%%%%%%%%%%%%%%%%%%%%%%%%%%%%%%%%%%%%%%%
\section{Monolithic Model}
\label{sec:Monolithic}
%%%%%%%%%%%%%%%%%%%%%%%%%%%%%%%%%%%%%%%%%%%%%%%%%%%%%%%%%%%%%%%%%%%%%%%%%%%%%%
We have, for completeness, trained a monolithic model based on fully connected neural networks. In this case, instead of performing a preliminary dimensional reduction before the final interpolation in $\omega/\mu_V$, we allow a single neural network to receive $\omega/\mu_V$ and produce $\psi_4^i$ directly. In this case, the real and imaginary part of the waveform are concatenated in a single vector of dimension $2N_\psi = 1024$, as complex values are not trivially handled by common activation functions. Cubic splines are used to learn the dependence of the peak position and amplitudes $(t_0, A)$, as defined in Section \ref{subsec:pre-processing}.

The wave-fitting network has an input layer with 1 neuron (receiving $\omega/\mu_V$) and successive layers of sizes 64, 256, 512 and 1024 (the output layer). We use GELU activation functions after each layer except the last one.

\begin{figure}[t]
\begin{center}
\includegraphics[width=0.45\textwidth]{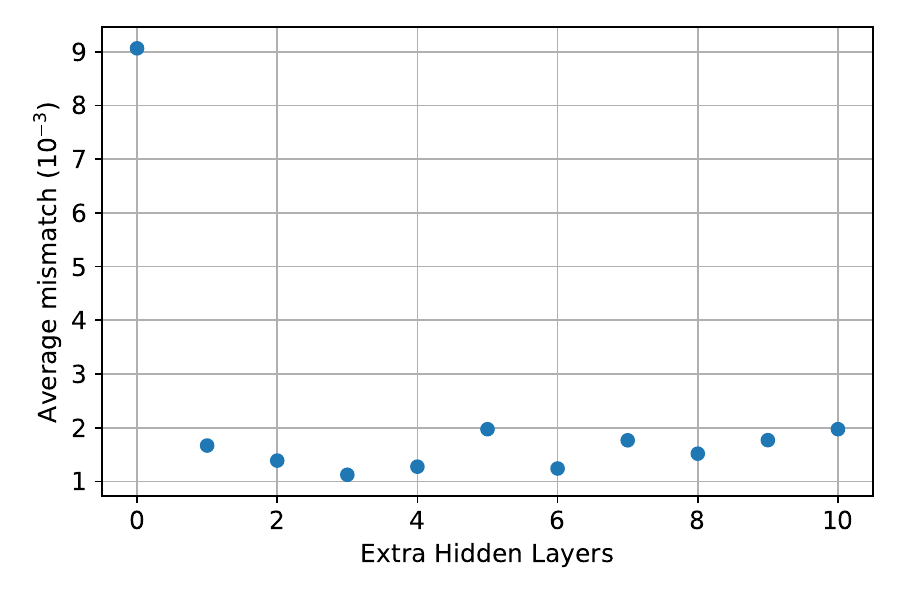}
\caption{Average mismatch of the monolithic model on the $k$-Fold cross-validation of the test set, for the mode $(l,m) = (2,2)$. The performance does not seem to vary as long as we include at least one extra hidden layer of 64 neurons.  \label{fig:Extra_Layers}}
\end{center}
\end{figure}

As in the case of the autoencoder, training is performed on a Nvidia RTX 4090 24GB GPU over 500 iteration epochs divided in minibatches of 16 waveforms, with an Adam optimizer of learning rate of $10^{-3}$, and $(\beta_1, \beta_2) = (0.5, 0.999)$.

In order to gauge the dependence of the model on hyperparameters, we introduce a number of replicas of the layer of size 64, exclusively on the wave-fitting network. As an example, the network with 2 extra hidden layers would have sizes $(1, 64, 64, 64, 256, 512, 1024)$. In Fig.~\ref{fig:Extra_Layers} we observe that the results improve significantly when we add one extra layer, but plateau after that. For this reason, we used one extra layer for the final model.
%
%%%%%%%%%%%%%%%%%%%%%%%%%%%%%%%%%%%%%%%%%%%%%%%%%%%%%%%%%%%%%%%%%%%%%%%%%%%%%%
\section{Results}
\label{sec:Results}
%%%%%%%%%%%%%%%%%%%%%%%%%%%%%%%%%%%%%%%%%%%%%%%%%%%%%%%%%%%%%%%%%%%%%%%%%%%%%%

\subsection{Model Evaluation}

\begin{figure}[thpb]
\begin{center}
\includegraphics[width=0.45\textwidth]{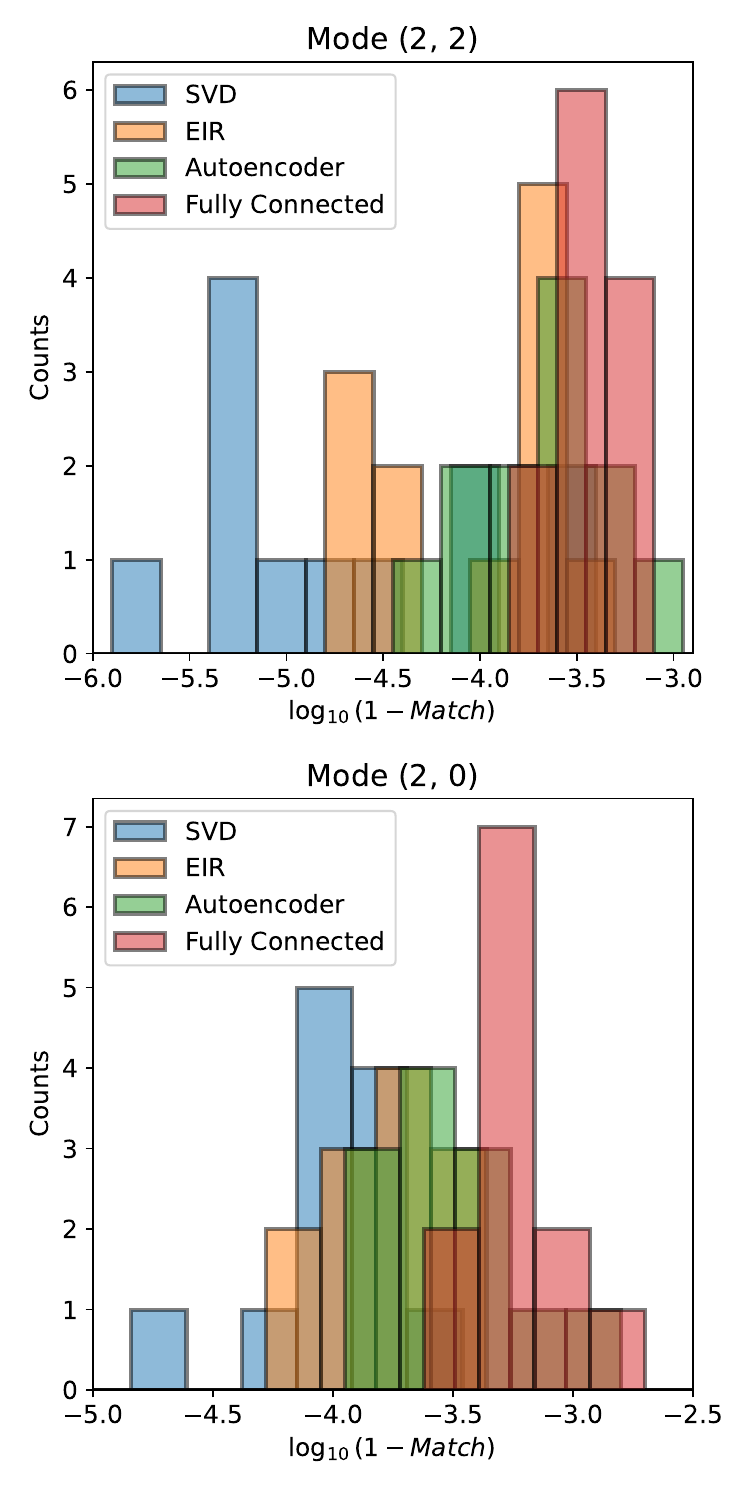}
\caption{Histograms of the distribution of the mismatches on the final test set, for the four tested model architectures. Lower mismatches correspond to a better model accuracy. We consider the two dominant modes: $(l, m) = (2, 2)$ in the upper plot, and  $(l, m) = (2, 0)$ in the lower plot. \label{fig:Model_Comparison_Hist}}
\end{center}
\end{figure}

Figure \ref{fig:Model_Comparison_Hist} shows histograms of the mismatches obtained on the final testing set by the four categories of models. We consider the two dominant modes, namely $(l,m) = (2,2)$ and $(l,m) = (2,0)$. In both cases we can observe that the SVD model seems to work best, followed by EIR, the autoencoder, and finally the monolithic fully-connected network. There is, however, a significant overlap between the distributions.

There are several possible reasons for the better performance of SVD. A lower number of parameters makes the model less prone to overfitting, and therefore more robust. Also, SVD is based on standard linear algebra algorithms, which do not require training by gradient descent, which removes the possible accuracy losses due to early or late stopping of the training process.

The significantly better performance of the SVD model on the testing set, together with its simplicity and the fact that its training is much faster than those based on neural networks, clearly indicate that this type of model is the most convenient for our physical system. Indeed, the parameter estimation Bayesian parameter inferences in Sections \ref{subsec:injections} and \ref{subsec:GW190521} have been performed with SVD models with 10 basis vectors and $N_\psi = 3001$, in this case trained with the complete dataset of 59 waveforms. 

\begin{figure*}[thpb]
\begin{center}
\includegraphics[width=\textwidth]{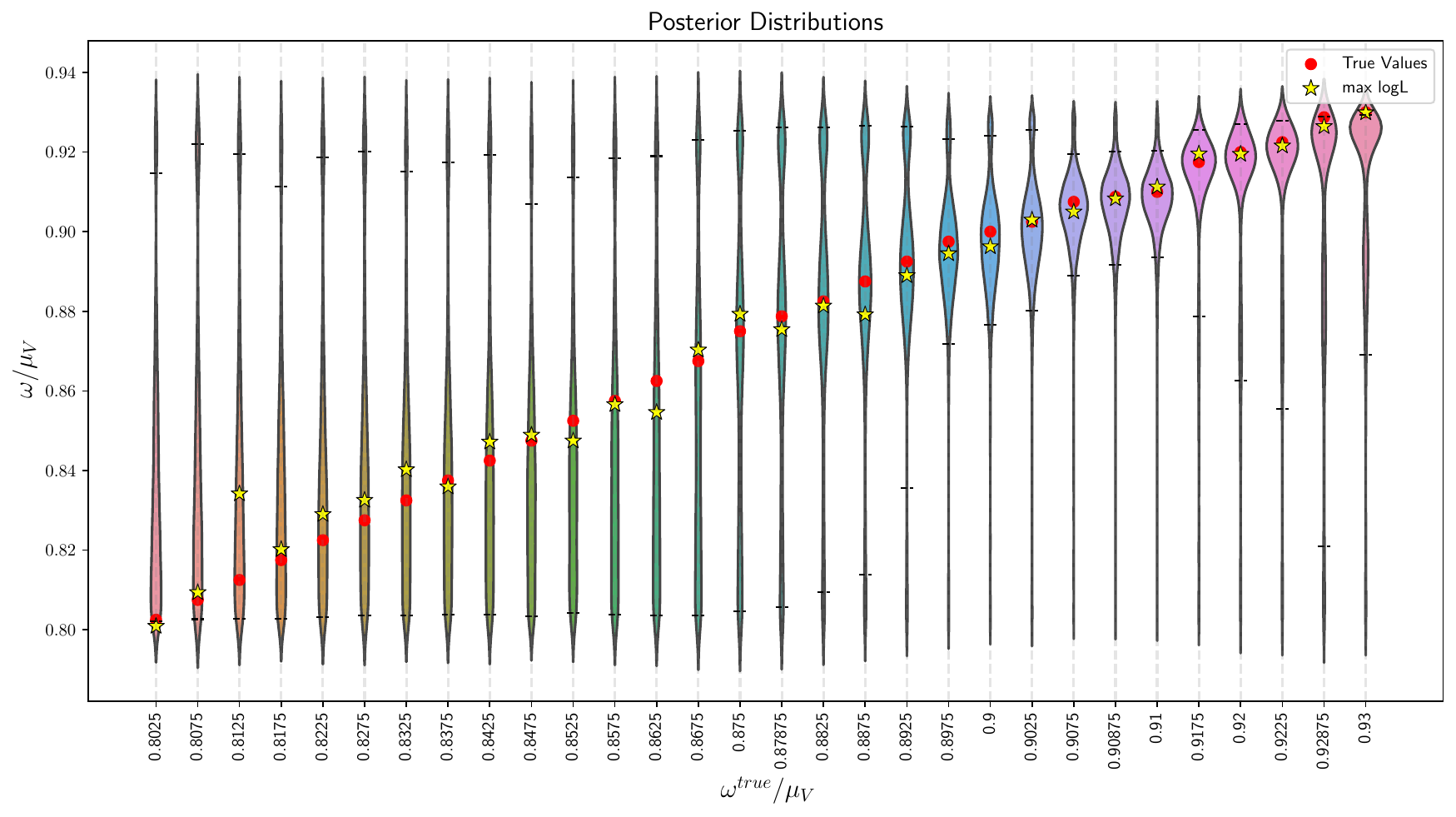}
\caption{Posterior distributions for the field frequency $\omega/\mu_V$ as a function of the corresponding true values. We obtain these through the injection of Numerical Relativity waveforms in zero-noise and their posterior recovery using our surrogate model. True values are denoted by red dots (red line) in the lower (upper) plot while maximum likelihood values are denoted by yellow stars. The horizontal bars delimit $90\%$ credible intervals.}
\label{fig:violin_omega}
\end{center}
\end{figure*}

\begin{figure*}[thpb]
\begin{center}
\includegraphics[width=\textwidth]{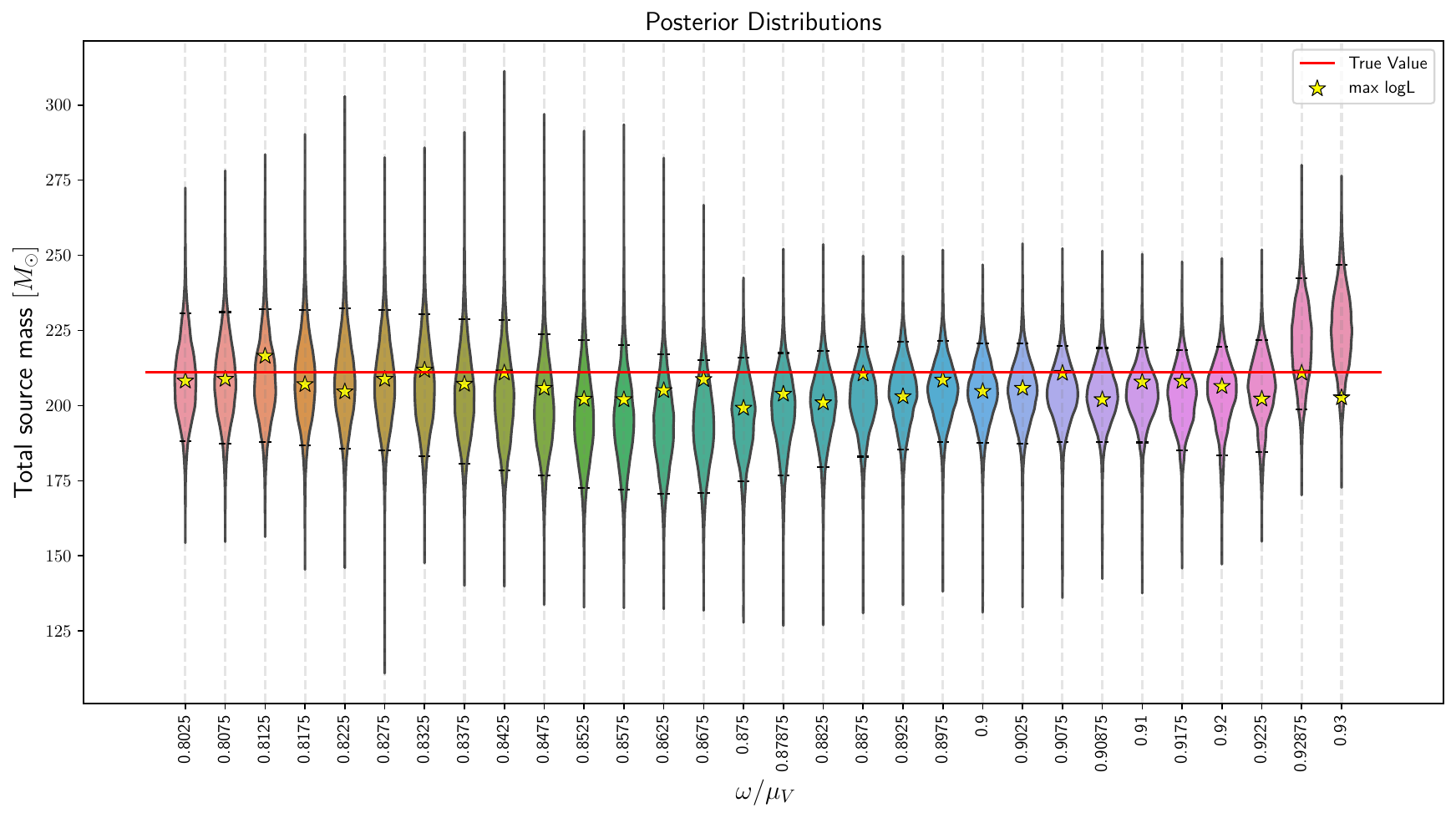}
\includegraphics[width=\textwidth]{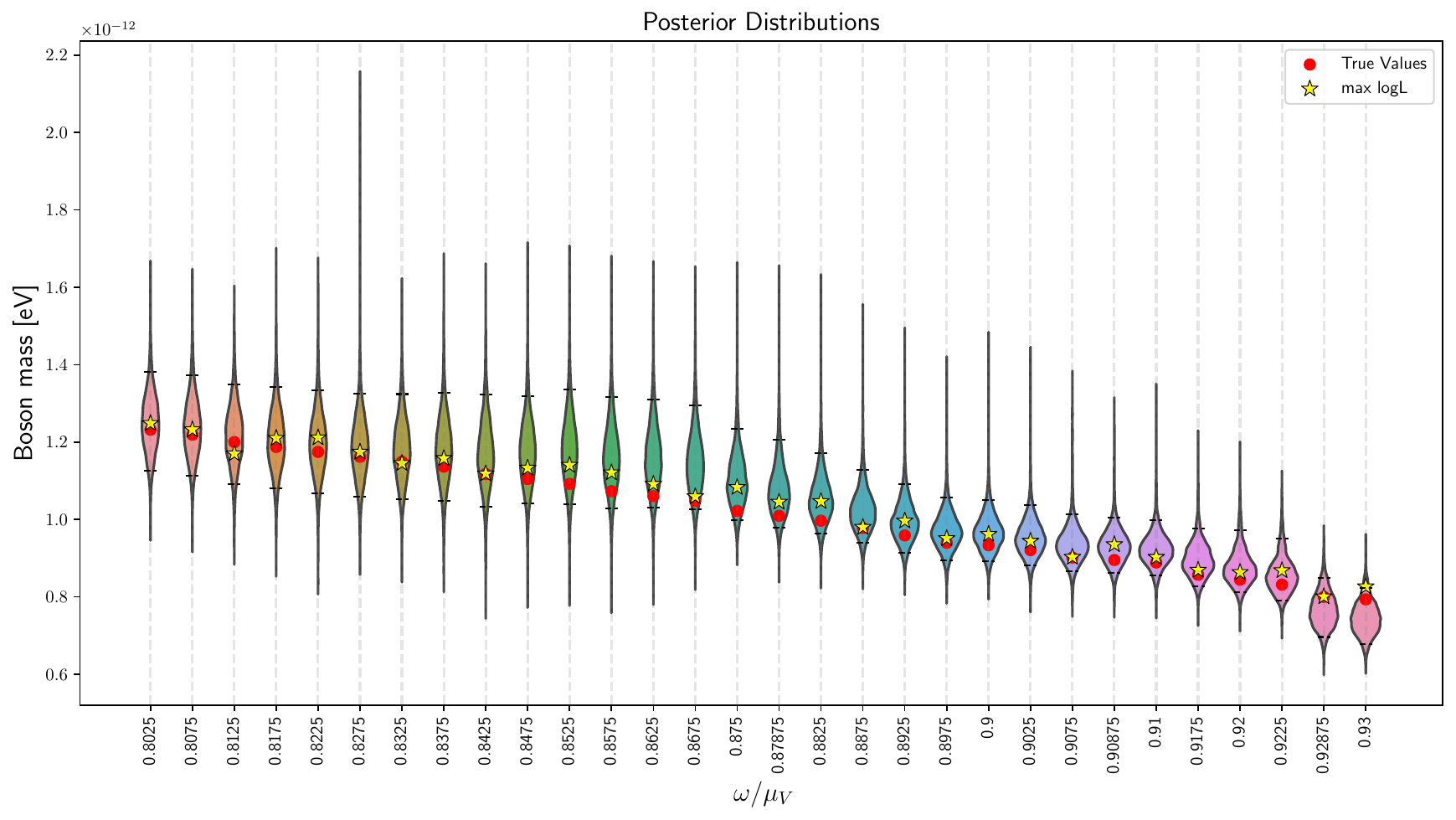}
\caption{Posterior distributions for the source-frame total mass (top) and the ultralight boson mass (bottom) as a function of the value of $\omega/\mu_V$. True values are denoted by red dots in the plot while maximum likelihood values are denoted by yellow stars. The horizontal bars delimit $90\%$ credible intervals.}
\label{fig:violin_mass}.
\end{center}
\end{figure*}

\subsection{Injection Recovery}
\label{subsec:injections}

We test the performance of our model in parameter inference and model selection tasks. To this end, we inject in zero noise a family of 30 signals from head-on Proca-star mergers of varying $\omega/\mu_V$, directly obtained through numerical-relativity simulations. We recover these injections using our SVD surrogate model together with the parameter inference code \texttt{Bilby} \cite{Ashton:2018jfp} in its parallelizable version \texttt{Parallel Bilby} \cite{pbilby_paper}. We set uniform priors in the detector-frame total mass and $\omega/\mu_V$. In addition, we set isotropic priors both on sky-location and source orientation, together with a uniform prior in the polarisation angle and a prior in distance uniform in co-moving volume. These priors match those used for the analyses presented in \cite{CalderonBustillo:2020fyi} and \cite{CalderonBustillo:2022dph} restricted equal-mass systems. We choose a three-detector network formed by the two Advanced LIGO detectors and Advanced Virgo equipped with their PSD at the time of the GW190521 event. We perform our analysis directly on the Newman-Penrose scalar $\psi_4$, as described in \cite{CalderonBustillo:2022dph}. We sample the parameter space using the sampler \texttt{Dynesty} \cite{Dynesty} with $N=1024$ live points. \\

Figure \ref{fig:violin_omega} shows the posterior distributions, together with $90\%$ credible intervals, for the field frequency $\omega/\mu_V$, as a function of the true injected value $\omega_{\rm true}/\mu_V$. The true values are denoted by red circles while maximum likelihood (ML) values (i.e.~those producing the best fits) are denoted by yellow stars. As somewhat expected from our match analyses, we find that the posterior distribution always peaks very close to the true value, with the latter being always contained within the $90\%$ credible intervals. The same is found in most cases for the ML values, which always lay very close to the true parameters \textit{found by the sampler}. We note that the sampler is not designed to find the actual best-fit parameters but to reconstruct the corresponding posterior probability distributions. Due to this, in some cases, the recovered ML parameters may not match the true ones, even if the waveform model is infinitely accurate. This is the case, for instance, when there are large degeneracies in the parameter space or when the sampling is not extremely aggressive. Nevertheless, since in certain situations parameter recovery can be strongly influenced by Bayesian priors causing shifts in the posteriors, we report the location of the ML points to show that such shifts are indeed sourced by priors and not by systematic errors due to the waveform model. 

We find that certain values of $\omega_{\rm true}$ lead to varying features in the posteriors. First, some cases lead to wide distributions while others lead to significantly sharper ones. This reflects the fact that, in certain regions of the parameter space, small changes in $\omega_{\rm true}$ cause significant changes in the waveform morphology while smoother changes are induced in other regions. Second, for some cases we find somewhat bi-modal distributions. Far from denoting any issues in the analysis or systematic errors in the waveform model, these features come as a consequence of the physics governing Proca-star waveforms. In particular, mergers corresponding to the lower and upper end of our $\omega/\mu_V$ range, are significantly louder than those corresponding to intermediate frequencies. Therefore, such configurations are highly favoured by our distance prior, producing secondary peaks of the posterior distribution either at large frequencies (for the lowest frequency injections), both large and low frequencies (for the mid-frequency injections), and almost imperceptible peaks at low frequencies for the highest-frequency injections. For the latter, the posterior also somewhat extends artificially to lower frequencies due to the upper frequency cutoff imposed by the model, which prevents the posterior distribution to extend beyond such cutoff.

Figure \ref{fig:violin_mass} shows the posterior distributions for the total mass $M_{\rm total}$ of the binary and the corresponding boson mass $\mu_V$. First, we note that the true values are again always contained within the $90\%$ credible intervals. Second, we note that the posteriors for $M_{\rm total}$ are systematically shifted towards lower values. Again, far from reflecting systematic errors in our model, this is due to the impact of the prior for the luminosity distance \cite{CalderonBustillo:2020fyi, CalderonBustillo:2022dph}. Recall that the source-frame total mass is $M_{\rm total}=M_{\rm det}/(1+z)$, where $M_{\rm det}$ and $z$ are, respectively, the detector-frame total mass and the redshift. Since the distance prior strongly favours large distances, we find that such estimates, and therefore those for $z$, are always shifted towards large values, causing the opposite effect on $M_{\rm total}$. The same effect is present but less pronounced for $\mu_V$, owing to its strong dependency on the -- better captured -- value of $\omega/\mu_V$. Finally, we note that ML values are always closer (when not directly on top) to the true injected values than the bulk of the posterior distributions.

\subsection{GW190521 Parameter Estimation}
\label{subsec:GW190521}

We perform parameter inference on the GW event GW190521, which has been previously analysed directly using numerical relativity simulations for head-on Proca-star waveforms, both restricted to equal-mass cases \cite{CalderonBustillo:2020fyi, CalderonBustillo:2022dph} and extended to unequal-mass ones \cite{CalderonBustillo:2022cja}. We set the exact same priors described in the previous section, which match those used in \cite{CalderonBustillo:2020fyi, CalderonBustillo:2022dph}. We will compare our results to those reported in the second column of Table III in \cite{CalderonBustillo:2022dph}, which unlike those in \cite{CalderonBustillo:2020fyi}, make direct use of the Newman-Penrose scalar. The two main differences between these two analyses are: a) The usage of a surrogate model instead of numerical relativity waveforms, and b) the continuous sampling of the parameter space as opposed to the exploration of a discrete set of values of $\omega/\mu_V$.

\begin{figure}[thpb]
\begin{center}
\includegraphics[width=0.45\textwidth]{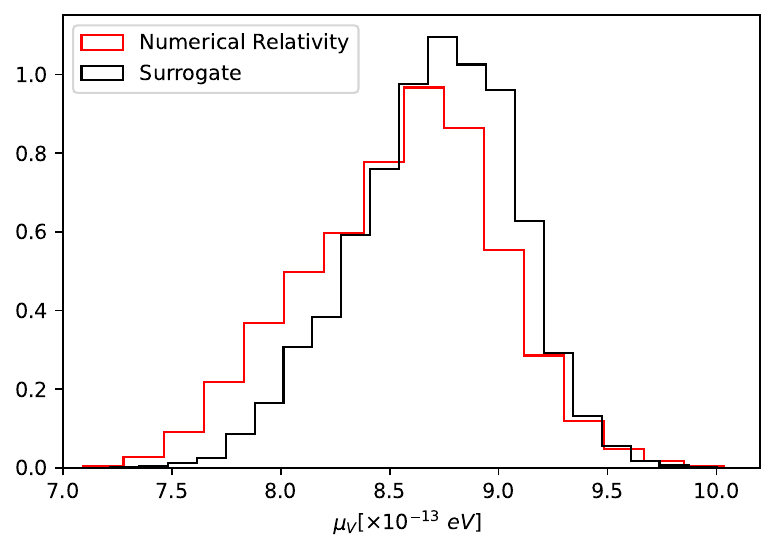}
\includegraphics[width=0.45\textwidth]{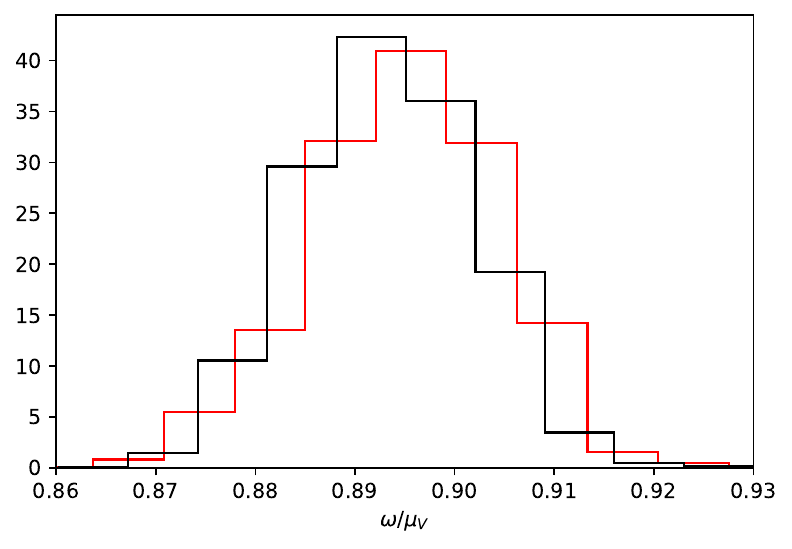}
\caption{Posterior probability distributions for the field frequency $\omega/\mu_V$ and the ultralight boson mass $\mu_V$ for GW190521. The black curve corresponds to the analysis carried out with our surrogate model. The red curve corresponds to the analysis carried out directly with numerical relativity templates, using the exact framework described in \cite{CalderonBustillo:2022dph,CalderonBustillo:2022cja}.}
\label{fig:PE_Posterior_Placeholder}
\end{center}
\end{figure}

Table \ref{tab:pe} reports the parameter estimates obtained from our analysis together with those obtained in \cite{CalderonBustillo:2022dph}, expressed as median values together with symmetric $90\%$ credible intervals. In addition, Figure \ref{fig:PE_Posterior_Placeholder} shows the corresponding posterior distributions for the field frequency $\omega/\mu_V$ and the ultralight-boson mass $\mu_V$. We note that the results are in very good agreement, indicating again the reliability of our surrogate model. Table \ref{tab:pe} also shows the corresponding natural log-Bayes' Factors with respect to the noise hypothesis, which match within the sampling uncertainty of 0.1. Finally, we find that the numerical relativity model recovers a slightly larger log likelihood, indicating that the best-fitting waveform fits the data marginally better than in the Surrogate case. We attribute this to the following fact. While for the case of the surrogate model we perform a single inference run over all waveform parameters, for the NR model we combine the result of many runs that sample over the extrinsic waveform parameters and total mass for fixed $\omega/\mu_{V}$ \cite{CalderonBustillo:2020fyi}, which leads to a more aggressive sampling of the parameter space. 

\begin{table}[t!]
\centering
\begin{center}
%\begin{ruledtabular}
%\begin{tabularx}{\columnwidth}{>{\raggedright\arraybackslash}Xrr}
\renewcommand{\arraystretch}{1.5}
\begin{tabular}{l @{\hspace{1em}}c@{\hspace{1em}} @{\hspace{1em}}c@{\hspace{1em}}}
%\hline
%\hline \\ 
Parameter  & Surrogate & NR  \\ \hline
Total / Final mass $[M_\odot]$ & $233^{+10}_{-12}$ & $233^{+12}_{-16}$
\\
Inclination [rad] & $0.84^{+0.21}_{-0.34}$ & $0.85^{+0.23}_{-0.36}$
\\
Luminosity distance [Mpc] & $549^{+240}_{-149}$  & $544^{+296}_{-163}$ 
\\
Redshift $z$ & $0.11^{+0.03}_{{-0.05}}$ & $0.11^{+0.03}_{{-0.05}}$
\\
Total red-shifted mass $[M_\odot]$  & $259^{+7}_{-8}$ & $260^{+8}_{-9}$ 
\\
Field frequency $\omega/\mu_{\rm V}$  & $0.893^{+0.014}_{-0.014}$ & $0.890^{+0.015}_{-0.015}$
\\
Boson mass $\mu_{\rm V}$ [$\times 10^{-13}$\,eV] & $8.73^{+0.53}_{-0.68}$ & $8.60^{+0.63}_{-0.62}$
\\
$\ln$ Bayes Factor S/N& $90.46$ & $90.36$ \\
$\ln$ Max. Likelihood & $116.1$ & $116.5$

%ULogBayesFactor^{\text{Proca}}_{\text{BBH,Best}}  & $3.1$ (NRSur) & $2.5$ (NRSur)  & $1.0$ (XPHM)  & ---
%Evidence for $(2,0)$ mode & $\log{\cal B}  \simeq 0.6$ & ---
%\\[6pt]
%\hline
%\hline
%\end{tabularx}
\end{tabular}
%\end{ruledtabular}
\caption{Parameters for GW190521 obtained under the  head-on Proca-star merger analysis using our surrogate model (middle column) and the approach based on numerical relativity waveforms presented in \cite{CalderonBustillo:2022dph} (right column).  We quote median values with symmetric $90\%$ credible intervals. The last two rows reports the natural logarithmic Bayes Factor for the signal vs.~noise hypotheses and the corresponding maximum likelihoods.}  
\label{tab:pe}
\end{center}
\end{table}

%%%%%%%%%%%%%%%%%%%%%%%%%%%%%%%%%%%%%%%%%%%%%%%%%%%%%%%%%%%%%%%%%%%%%%%%%%%%%%
\section{Discussion}
\label{sec:Discussion}
%%%%%%%%%%%%%%%%%%%%%%%%%%%%%%%%%%%%%%%%%%%%%%%%%%%%%%%%%%%%%%%%%%%%%%%%%%%%%%

We have constructed several accurate surrogate models that generate waveforms of head-on collisions of equal-mass Proca stars. The models are purely data-driven, without any predefined physical input, and are trained on a catalog of 59 numerical relativity waveforms with several multipoles each. Different model architectures have been analyzed. On the one hand, a collection of two-stage models, which implement a dimensional reduction algorithm followed by cubic spline interpolation of the governing parameters. The dimensional reduction algorithms discussed comprise singular value decomposition, empirical interpolant representation, and finally a deep convolutional autoencoder. On the other hand, we have also tested a monolithic model where a fully-connected neural network interpolates the full waveform time series as a function of the physical parameter $\omega/\mu_V$.

While all four models have proven to be accurate, with all mismatches below $10^{-3}$ for the mode $(l,m) = (2,2)$, the model based on singular value decomposition has proven to be the most precise. In terms of precision, SVD is followed by the empirical interpolant representation, the autoencoder and finally the monolithic fully-connected network. The two first models are not only the most precise but also much faster than those based on neural networks, as they do not require gradient descent optimization. However, the neural networks are able to encode waveforms nonlinearly from very few input real parameters (2 for the autoencoder, 1 for the fully-connected), while SVD and EIR require at least 7 complex parameters for our dataset.

We have tested the SVD surrogate model within the task of Bayesian parameter inference through the recovery of numerical relativity injections in zero noise, obtaining excellent results. Finally, we have also performed parameter estimation on GW190521 data, finding strong agreement with the results obtained directly using numerical relativity templates. As a general summary, we conclude that the type of surrogate models investigated in this work can be regarded as a convenient tool to efficiently perform Bayesian analysis on gravitational waveforms from bosonic star mergers. 

A natural extension of this work is the inclusion of waveforms from Proca star mergers with unequal masses. Our early attempts at this extension, however, have encountered serious difficulties due to the interference effects between the stars at merger~\cite{Sanchis-Gual:2022mkk}. When the Proca fields from the two colliding stars oscillate at different frequencies $\omega_1$ and $\omega_2$, the relative phase at the point of encounter depends on the particular values of these frequencies. This causes an interference pattern, where the amplitude of the emitted gravitational waves presents strong variations. Additionally, at the regions of lowest amplitude, the shape of the waveform changes very rapidly with $\omega_1$ and $\omega_2$.  At present, our catalog of simulations is not sufficiently dense in these difficult regions for the model to reach the desired precision. 

%%%%%%%%%%%%%%%%%%%%%%%%%%%%%%%%%%%%%%%%%%%%%%%%%%%%%%%%%%%%%%%%%%%%%%%%%%%%%%
\begin{acknowledgments}
%%%%%%%%%%%%%%%%%%%%%%%%%%%%%%%%%%%%%%%%%%%%%%%%%%%%%%%%%%%%%%%%%%%%%%%%%%%%%%

%
We would like to thank Koustav Chandra for his very useful comments on the first manuscript.
RL acknowledges financial support provided by Generalitat Valenciana / Fons Social Europeu through APOSTD 2022 post-doctoral grant CIAPOS/2021/150.
JCB is supported by a fellowship from ``la Caixa'' Foundation (ID
100010434) and from the European Union’s Horizon 2020 research and innovation programme under the Marie Skłodowska-Curie grant agreement No 847648. The fellowship code is LCF/BQ/PI20/11760016. JCB is also supported by the research grant PID2020-118635GB-I00 from the Spain-Ministerio de Ciencia e Innovaci\'{o}n.
NSG acknowledges support from the Spanish Ministry of Science and Innovation via the Ram\'on y Cajal programme (grant RYC2022-037424-I), funded by MCIN/AEI/10.13039/501100011033 and by ``ESF Investing in your future”.
MLl-M, NSG, ATF and JAF are supported by the Spanish Agencia Estatal de Investigaci\'on (grant PID2021-125485NB-C21) funded by MCIN/AEI/10.13039/501100011033 and ERDF A way of making Europe, and the Generalitat Valenciana (grant CIPROM/2022/49). We acknowledge further support by the European Horizon Europe staff exchange (SE) programme HORIZON-MSCA2021-SE-01 Grant No. NewFunFiCO-101086251.
This work is supported by the Center for Research and Development in Mathematics and Applications (CIDMA) through the Portuguese Foundation for Science and Technology (FCT -- Fundaç\~ao para a Ci\^encia e a Tecnologia), https://doi.org/10.54499/UIDB/04106/2020 and https://doi.org/10.54499/UIDP/04106/2020. The authors also  acknowledge support from the projects http://doi.org/10.54499/PTDC/FISAST/3041/2020,  http://doi.org/10.54499/CERN/FIS-PAR/0024/2021 and https://doi.org/10.54499/2022.04560.PTDC.
The authors acknowledge computational resources provided by the CIT cluster of the LIGO Laboratory and supported by National Science Foundation Grants PHY-0757058 and PHY0823459; and the support of the NSF CIT cluster for the provision of computational resources for our parameter inference runs. This material is based upon work supported by NSF's LIGO Laboratory which is a major facility fully funded by the National Science Foundation. The analysed LIGO-Virgo data and the corresponding power spectral densities, in their strain versions, are publicly available at the online Gravitational-Wave Open Science Center~\cite{SoftwareX,OpenDataArxiv}. This research has made use of data or software obtained from the Gravitational Wave Open Science Center (gwosc.org), a service of LIGO Laboratory, the LIGO Scientific Collaboration, the Virgo Collaboration, and KAGRA. LIGO Laboratory and Advanced LIGO are funded by the United States National Science Foundation (NSF) as well as the Science and Technology Facilities Council (STFC) of the United Kingdom, the Max-Planck-Society (MPS), and the State of Niedersachsen/Germany for support of the construction of Advanced LIGO and construction and operation of the GEO600 detector. Additional support for Advanced LIGO was provided by the Australian Research Council. Virgo is funded, through the European Gravitational Observatory (EGO), by the French Centre National de Recherche Scientifique (CNRS), the Italian Istituto Nazionale di Fisica Nucleare (INFN) and the Dutch Nikhef, with contributions by institutions from Belgium, Germany, Greece, Hungary, Ireland, Japan, Monaco, Poland, Portugal, Spain. KAGRA is supported by Ministry of Education, Culture, Sports, Science and Technology (MEXT), Japan Society for the Promotion of Science (JSPS) in Japan; National Research Foundation (NRF) and Ministry of Science and ICT (MSIT) in Korea; Academia Sinica (AS) and National Science and Technology Council (NSTC) in Taiwan. This manuscript has LIGO DCC number P2400107.

%%%%%%%%%%%%%%%%%%%%%%%%%%%%%%%%%%%%%%%%%%%%%%%%%%%%%%%%%%%%%%%%%%%%%%%%%%%%%%
\end{acknowledgments}
%%%%%%%%%%%%%%%%%%%%%%%%%%%%%%%%%%%%%%%%%%%%%%%%%%%%%%%%%%%%%%%%%%%%%%%%%%%%%%

\bibliography{references}

\end{document}